# Methylation Operation Wizard (MeOW): Identification of differentially methylated regions in long-read sequencing data

Miranda PG Zalusky[1,*], Danny E Miller[1,2,3,*]

[1] Department of Pediatrics, University of Washington, Seattle, WA 98195, USA
[2] Department of Laboratory Medicine and Pathology, University of Washington, Seattle, WA 98195, USA
[3] Brotman Baty Institute for Precision Medicine, University of Washington, Seattle, WA 98195, USA

**\*Correspondance**: mgaley@uw.edu, dm1@uw.edu

**Keywords**: Long-read sequencing, Nanopore sequencing, PacBio sequencing, differential methylation, methylation analysis

## ABSTRACT

Long-read sequencing (LRS) is able to simultaneously capture information about both DNA sequence and modifications, such as CpG methylation in a single sequencing experiment. Here we present Methylation Operation Wizard (MeOW), a program to identify and prioritize differentially methylated regions (DMRs) genome-wide using LRS data. MeOW can be run using either a file containing counts of per-nucleotide methylated CpG sites or with a bam file containing modified base tags.

## Availability

MeOW is freely available on GitHub at https://github.com/mgaleyuw/MeOW

## Contact

mgaley@uw.edu, dm1@uw.edu

## INTRODUCTION

Because more than half of individuals with a suspected Mendelian disorder remain undiagnosed after a complete evaluation, there is broad interest in using new technologies to evaluate individuals who are unsolved following a standard clinical genetic evaluation. Technologies such as long-read sequencing (LRS) have shown promise in identifying disease-causing variants that may be overlooked or difficult to detect using prior testing modalities (Wojcik *et al.*, 2023). LRS is unique in that as a single data source it can identify single nucleotide variants (SNVs), insertion or deletion variants (indels), structural variants (SVs), and differences in methylation (Logsdon *et al.*, 2020; Mastrorosa *et al.*, 2023). While many tools exist for the identification and analysis of these variant types, there are few capable of performing genome-wide evaluation for changes in methylation – especially at defined CpG sites – that may help explain the phenotype of an individual with a suspected Mendelian condition.

To address this need, we developed Methylation Operation Wizard (MeOW), software that can identify differentially methylated regions (DMRs) within individual samples at user-defined regions using a control database as reference points **(Figure 1A)**. MeOW runs quickly, taking approximately 20 minutes on a modestly powered computer to complete analysis on a human genome at 40x depth of coverage. Here, we show that MeOW is able to identify a clinically relevant DMR in an individual with Prader-Willi syndrome, a known imprinting disorder. MeOW is robust to differences in tissue type between reference sequences and the query set when the beta regression test parameter is used, but it can also be used to identify DMRs associated with differences in tissue type if lower thresholds and more permissive parameter settings are used. MeOW will simplify genome-wide analysis of challenging unsolved cases and enable identification of novel DMRs associated with human disease.

## METHODS

Meow is implemented in Rust 1.72.1, Python 3.9.6 and R 4.2.1. We developed and tested MeOW using publicly available data from the 1000 Genomes ONT Sequencing Consortium (see Data Availability). Briefly, DNA for those samples was isolated from pelletted lymphocyte cells using a modified Gentra Puregene protocol. The ligation kit SQK-LSK110 was used to prepare ONT libraries, which were sequenced on the ONT PromethION platform using R9.4.1 flowcells. DNA for the query case (M0168) was prepared using DNA extracted from 2 mL of fresh blood using a modified Gentra Puregene protocol, and sequencing libraries were prepared using the SQK-LSK110 kit and sequenced on an R9.4.1 flowcell on the ONT PromethION to an average depth of coverage of 40x. All data were basecalled using Guppy version 6.3.2 and 6.4.6 (ONT) using the super accurate model with 5mCG detection and aligned to GRCh38 using minimap2 (version 2.24) (Li, 2018).

MeOW processes bam files with 5mCG methylation values stored in the 'ML' and 'MM' tags into pileups at CpG positions defined by the UCSC genome browser (last accessed 2020-07-02) using samtools version 1.17 (Li *et al.*, 2009). Reads with mapping quality less than 1, individual bases with quality scores less than 1, and positions with insertions or deletion variants are excluded. MeOW calculates the weighted mean 5mCG methylation frequency for modified cytosines detected directly and modified cytosines detected via the complementary guanine on the forward strand. The probability of methylation is rescaled from a [1,255] interval to [0,1] and weighted according to the ML probability of methylation at each read reported in the pileup.

Input data sources are user-defined. A typical use of MeOW is that a control set is designated (which may be a cohort of unaffected individuals or samples) and then joined in a reference database, where positions missing data in any sample are discarded for all samples. Remaining CpG positions are used to create an index file with gene annotations from Ensembl (release 105) and a unique identifier for each CpG island defined by the UCSC genome browser (also implemented in Rust 1.72.1) (Kent *et al.*, 2002; Martin *et al.*, 2023). Methylation frequencies are shuffled across positions when the control database is made. It is recommended that at least 10 samples are used for a control database. The list of 1000 Genomes samples that were used to create the reference database is available on GitHub. The input data that is designed as the test case is indexed to the control database.

In R, the methylation frequency of the test sample is compared systematically to the methylation frequencies of the control data using paired *t*-tests with corrections for multiple hypotheses and beta regression. To account for variability in methylation frequency of control samples while maintaining information that may be contained in the positional sequence of CpGs, the *t*-test is bootstrapped a number of times set by the user (default is 50) and the average result is taken with variance reported. Beta regression tests are similarly run on shuffled data. Additionally, to reduce the rate of false positives, Cohen's *d* is reported as a measure of effect size, where 1.75 is a recommended cutoff value. Tests of CpG islands are limited to regions containing at least 50 C or G positions by default, but can be changed by the user at runtime.

**EXAMPLE ANALYSIS**

We used a control set of whole-genome LRS data from 19 randomly selected samples sequenced by the 1000 Genomes ONT Sequencing Consortium to assemble a reference database of CpG methylation variation. The 19 control datasets were used with MeOW to produce a reference database of 21,296 CpG islands that contained 50 or more C or G positions. As a control test case, we used DNA from a deidentified residual sample from an individual with a known imprinting disorder, Prader-Willi syndrome (MIM: 176270), caused by heterodisomy (the individual inherited two different copies of chromosome 15 from their mother). In this case, we would expect both copies of chromosome 15 to be methylated at the CpG island at the *SNURF/SNRPN* locus.

Differential methylation in the test dataset (M0168) was assessed using the beta regression test statistic option using the default parameters of 50 test bootstraps, a *p*-value cutoff of 0.01, and a Cohen's *d* cutoff of 1.75. In total, five hypermethylated CpG islands were identified, including one located in *SNURF/SNRPN* (chr22:24,954,889–24,955,907). This CpG island (cpg.7735) had an associated *p*-value of 0.0007 and a Cohen's *d* magnitude effect size of 1.76 **(Figure 1B)**. Two other significantly differentially methylated CpG islands were found outside of mapped genes, and two others were in nuclear pore complex-interacting protein genes **(Figure 1C, Figure 1D)**. In all differentially methylated CpG islands except cpg.7735, the average methylation frequency in controls was less than 0.25 while the frequency in M0168 was above 0.7. In cpg.7735, control samples have an average of 0.5 methylation frequency, while the average in M0168 is above 0.8, suggesting that both haplotypes were methylated, which is consistent with the known mechanism in which both copies of chromosome 15 were maternally inherited.

**DISCUSSION**

Here, we present MeOW, a program that can perform genome-wide analysis for DMRs from LRS data. MeOW includes a leave-one-out analysis routine for validation and assessment of variation within control datasets. We show that MeOW can perform genome-wide analysis and identify increased methylation at a single CpG site known to be associated with an imprinting condition. A control database was generated using publicly available data from the 1000 Genomes ONT Sequencing Consortium, which is not ideal in this case as DNA from the test sample used was derived from blood while the DNA for the 1000 Genomes samples was derived from lymphoblastoid cell lines. Despite the different DNA sources, MeOW was able to correctly identify and prioritize the key CpG in the test case.

While MeOW is initially released with a control database built from publicly available data from the 1000 Genomes ONT Sequencing Consortium, users are encouraged to use their own curated datasets of reference samples created using samples from the same tissue type, sequencing chemistry, and base calling program as their query sample. Because MeOW will work with any bam file that stores methylation information in the MM and ML tags it can be used to analyze LRS data generated on both the Nanopore and PacBio platforms. We anticipate MeOW will lead to the identification of novel differentially methylated regions when applied to large cohorts of individuals with unsolved Mendelian conditions.

## ACKNOWLEDGEMENTS

We thank Angela Miller for help with manuscript and figure preparation and Sophia Gibson for testing and feedback.

## IRB APPROVAL

Use of residual samples for testing was approved by the Seattle Children's Hospital IRB, study #00003536.

## CONFLICT OF INTEREST

DEM is on a scientific advisory board at Oxford Nanopore Technologies (ONT), is engaged in a research agreement with ONT, and they have paid for his travel to speak on their behalf. DEM holds stock options in MyOme.

## FUNDING

DEM is supported by the National Institutes of Health through the NIH Director's Early Independence Award DP5OD033357.

## DATA AVAILABILITY

Data from the 1000 Genomes ONT Sequencing Consortium is available at https://s3.amazonaws.com/1000g-ont/index.html. Summary CpG data from the control sample is available upon request.

**FIGURE**

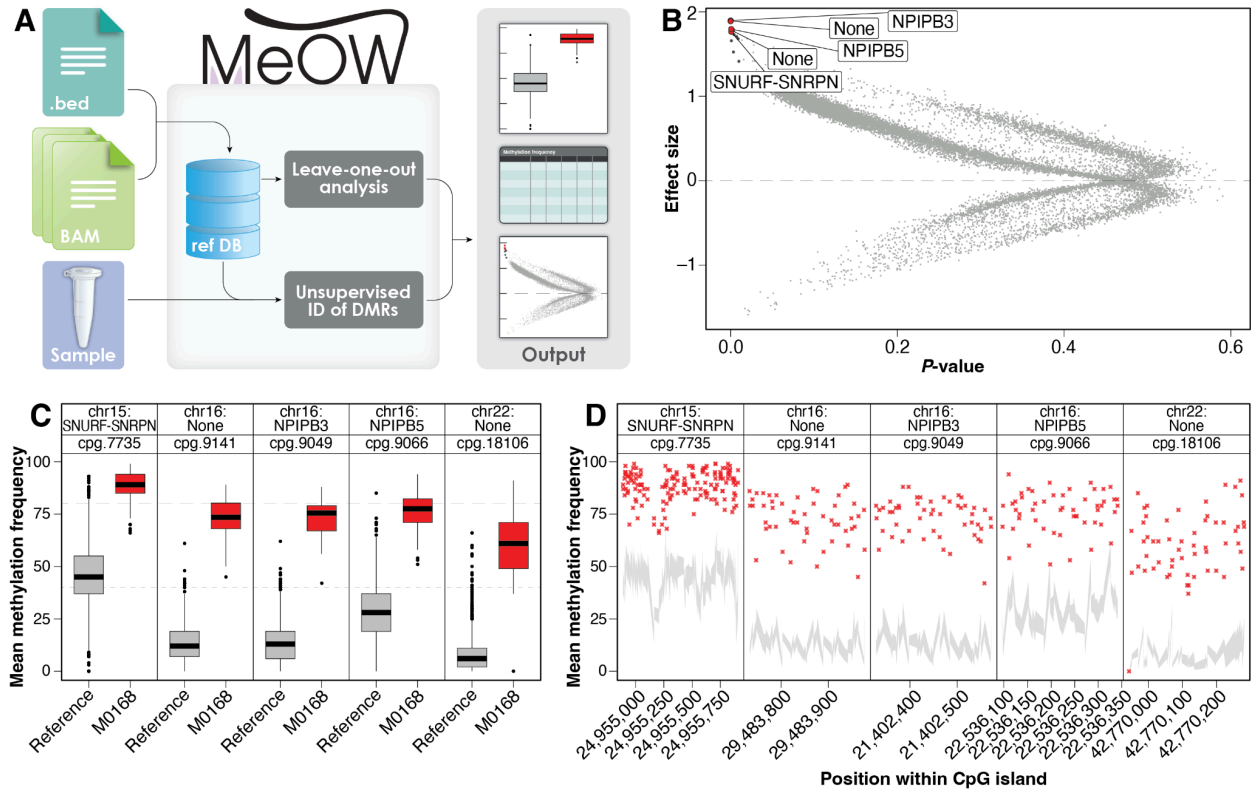

**Figure 1. MeOW identifies differentially methylated regions in long-read sequencing data.**
**A.** MeOW requires a set of aligned BAM files with the MM and ML tags populated as well as a bed file containing a list of regions of interest, such as CpG islands in order to build a reference database. After a reference database is built a leave-one-out analysis can be performed in the reference cohort to identify unique differentially methylated regions (DMRs) within that dataset. MeOW can also be run using a test sample against an already constructed reference database to identify DMRs. Output from both methods is available in table or graphical formats. **B.** Significant differentially methylated CpG sites are shown (red) for a test sample known to have Prader-Willi syndrome compared to a control database of 19 random samples sequenced as part of the 1000 Genomes Project ONT Sequencing Consortium. **C.** MeOW generates graphics illustrating the significant differences in methylation frequency between the test sample and the control dataset. The five DMRs shown represent the significant values from (B). **D.** Ribbon plots show the methylation frequency of each C and G in the query relative to the mean and standard error of control database methylation frequency at the same positions.

# REFERENCES

Kent,W.J. *et al.* (2002) The Human Genome Browser at UCSC. *Genome Res.*, **12**, 996–1006.

Li,H. (2018) Minimap2: pairwise alignment for nucleotide sequences. *Bioinformatics*, **34**, 3094–3100.

Li,H. *et al.* (2009) The Sequence Alignment/Map format and SAMtools. *Bioinformatics*, **25**, 2078–2079.

Logsdon,G.A. *et al.* (2020) Long-read human genome sequencing and its applications. *Nat Rev Genet*, **21**, 597–614.

Martin,F.J. *et al.* (2023) Ensembl 2023. *Nucleic Acids Research*, **51**, D933–D941.

Mastrorosa,F.K. *et al.* (2023) Applications of long-read sequencing to Mendelian genetics. *Genome Med*, **15**, 42.

Wojcik,M.H. *et al.* (2023) Beyond the exome: What's next in diagnostic testing for Mendelian conditions. *The American Journal of Human Genetics*, **110**, 1229–1248.